\documentclass{IEEEtran4PSCC}
\ifCLASSINFOpdf
   \usepackage[pdftex]{graphicx}
\else
   \usepackage[dvips]{graphicx}
\fi
\usepackage[cmex10]{amsmath}

\usepackage{todonotes} 

\usepackage{tabularx,adjustbox}
\usepackage{amsmath,amssymb,bm,mathrsfs}
\usepackage{siunitx}

\usepackage{breqn}
\usepackage{amsmath}
\usepackage{graphicx}
\usepackage{mwe}
\usepackage{arydshln}
\usepackage{makecell}
\usepackage{adjustbox}
\usepackage{float}
\usepackage{nomencl}
\makenomenclature
\usepackage{mathtools}
\usepackage{bm}
\usepackage{soul}
\usepackage{siunitx}
\usepackage{cite}
\usepackage{multirow}
\usepackage{arydshln}
\usepackage{xfrac}
\usepackage{nicefrac}
\usepackage{nicematrix,tikz}
\usepackage[T1]{fontenc}
\usepackage{inputenc}
\usepackage{babel}
\usepackage[font=small,labelfont=bf]{caption}
\DeclareSIUnit \var { var } 
\usepackage{titlesec}
\usepackage{etoolbox}
\usepackage{arydshln} 
\usepackage{environ}         
\usepackage{etoolbox}        
\usepackage{graphicx}        
\usepackage{hyperref}

\newlength{\myl}
\let\origequation=\equation
\let\origendequation=\endequation

\RenewEnviron{equation}{
  \settowidth{\myl}{$\BODY$}                       
  \origequation
  \ifdimcomp{\the\linewidth}{>}{\the\myl}
  {\ensuremath{\BODY}}                             
  {\resizebox{0.89\linewidth}{!}{\ensuremath{\BODY}}}  
  \origendequation
}
\newcommand{\overbar}[1]{\mkern 1.5mu\overline{\mkern-1.5mu#1\mkern-1.5mu}\mkern 1.5mu}

\def\hlinewd#1{%
\noalign{\ifnum0=`}\fi\hrule \@height #1 %
\futurelet\reserved@a\@xhline}
\makeatother

\makeatletter
\let\old@ps@headings\ps@headings
\let\old@ps@IEEEtitlepagestyle\ps@IEEEtitlepagestyle
\def\psccfooter#1{%
    \def\ps@headings{%
        \old@ps@headings%
        \def\@oddfoot{\strut\hfill#1\hfill\strut}%
        \def\@evenfoot{\strut\hfill#1\hfill\strut}%
    }%
    \def\ps@IEEEtitlepagestyle{%
        \old@ps@IEEEtitlepagestyle%
        \def\@oddfoot{\strut\hfill#1\hfill\strut}%
        \def\@evenfoot{\strut\hfill#1\hfill\strut}%
    }%
    \ps@headings%
}
\makeatother

\psccfooter{%
        \parbox{\textwidth}{\hrulefill \\ \small{23rd Power Systems Computation Conference} \hfill \begin{minipage}{0.2\textwidth}\centering \vspace*{4pt} \includegraphics[scale=0.06]{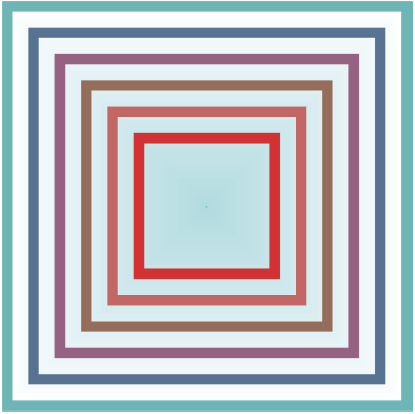}\\\small{PSCC 2024} \end{minipage} \hfill \small{Paris, France --- June 4 -- 7, 2024}}%
}

\begin{document}
%
\title{Experimental Validation of a Grid-Aware Optimal Control of Hybrid AC/DC Microgrids}

\author{\IEEEauthorblockN{Willem Lambrichts\IEEEauthorrefmark{1},
Jules Macé\IEEEauthorrefmark{2} and
Mario Paolone\IEEEauthorrefmark{1}}
\IEEEauthorblockA{\IEEEauthorrefmark{1} Distributed Electrical System Laboratory (DESL) \\
EPFL,
Lausanne, Switzerland\\ }
\IEEEauthorblockA{\IEEEauthorrefmark{2} Power Electronics Laboratory (PEL)\\
EPFL,
Lausanne, Switzerland\\ }
}


\maketitle

\begin{abstract}

This paper presents the experimental validation of a grid-aware real-time control method for hybrid AC/DC microgrids. The optimal control is leveraged by the voltage sensitivity coefficients (SC) that are computed analytically using the close-form expression proposed in the authors' previous work. The SCs are based on the unified power flow model for hybrid AC/DC grids that accounts for the AC grid, DC grid, and the Interfacing Converters (IC), which can operate in different control modes, e.g. voltage or power control. The SCs are used to express the grid constraints in the optimal control problem in a fully linear way and, therefore, allow for second- to subsecond control actions.
The validation of the model is performed on the hybrid AC/DC grid, available at the EPFL. The network consists of 18 AC nodes, 8 DC nodes, and 4 converters to interface the AC and DC network. The network hosts multiple controllable and uncontrollable resources.
The SC-based optimal control is validated in a generic experiment. It is shown that the real-time control is able to control the ICs optimally to redirect power through the DC grid, to avoid grid constraint violations while providing reactive power support to the upper layer AC grid. Furthermore, the computational time of the optimal control is analysed to validate its application in critical real-time applications.
\end{abstract}

\begin{IEEEkeywords}
Hybrid AC/DC networks, Sensitivity Coefficients, Optimal Control, Experimental Validation
\end{IEEEkeywords}

\thanksto{\noindent The project has received funding from the European Union’s Horizon 2020 Research \& Innovation Programme under grant agreement No. 957788.}

\section{Introduction}
\label{sec:intro}

Hybrid AC/DC microgrids are a promising solution for future power grids that are relying heavily on renewable sources. In fact, integrating AC and DC networks have several advantages, such as an increased overall efficiency of the system \cite{eghtedarpour2014power}, allows for more flexible control (through the presence of controllable AC/DC Interfacing Converters (IC)), and reduces the cost of the system because fewer power conversion sources are required as DC sources and loads are directly connected in the DC grid) \cite{BookACDCcontrol, eghtedarpour2014power}.

Real-time optimal control strategies are crucial for the operation of such grids to regulate the various Distributed Energy Resources (DER) in an optimal way in order to avoid, e.g., grid constraint violations while achieving a certain objective, such as minimising losses or maximising self-consumption.
Optimal Power Flow (OPF) in hybrid AC/DC networks has been intensively studied in the past, particularly in the application of High-Voltage Direct Current systems \cite{yang2017optimal, hotz2019hynet}. However, for AC/DC networks with multiterminal architectures, the literature has presented less robust solutions for optimal control strategies. In general, the proposed methodologies predominantly rely on three approaches: 1) droop control mechanisms of the IC \cite{eghtedarpour2014power, eajal2016unified, mevsanovic2018robust}, 2) decomposition of the problem where the AC and DC system are treated individually \cite{qachchachi2014optimal, hosseinzadeh2015robust} or 3) relaxation techniques of the non-convex ICs' constraints using second-order cone programming \cite{baradar2013second, li2018optimal}.


The nature of the above-mentioned problem formulations generally limits the versatility of the different control modes of ICs and, furthermore, restricts the regulation of the DC voltage to a single IC \cite{alvarez2021universal}. Generally, the inner control loops of the IC regulate two variables simultaneously as a result of the decoupling of the d and q frames. Typically, the DC voltage or active power is controlled together with the reactive power. The control modes are referred to as power control: $P_{ac} - Q_{ac}$ , or voltage control: $E_{dc} - Q_{ac}$.
The existing methods presented in the literature allow only one IC to regulate the DC voltage. Therefore, the flexibility of the hybrid network and the security of supply during, e.g. islanding manoeuvres, are greatly reduced. Indeed, when multiple voltage-controlled ICs are present, the power required to obtain the DC voltage can be shared over multiple ICs. This allows for a broader operation and improves the redundancy of the system in the event of failure. Furthermore, it also allows for different voltage levels within the DC grid, which makes it interesting for resources operating at different voltages \cite{barcelos2022direct}.

In view of the above, this paper presents an experimental validation of an optimal real-time control algorithm for hybrid AC/DC networks. The optimal control is based on the unified Power Flow (PF) method for hybrid grids presented in \cite{willem_PF}. The hybrid model includes the AC grid, DC grid, and ICs, which can operate on different control modes (voltage or power control) and, compared to other works presented in the literature, allows multiple ICs to regulate the DC voltage. 

The optimal control is based on the linearisation of the unified power flow model, usually referred to as sensitivity coefficients (SC). These are the partial derivatives of the nodal voltages and branch currents with respect to the nodal power injections. 
The SCs allow to formulate the non-convex voltage and current flow equations as a linear constraint in the optimal control problem. 

This work uses the method proposed in \cite{christakou2015real} to obtain a closed-form expression of the SC allowing an efficient analytical computation. The method is extended for hybrid AC/DC networks in \cite{willem_SC} and allows to include the unified grid model as constraints of the optimal control problem while accounting for the different control modes of the IC. 
The linearised OPF formulation is very well suited for real-time control where the network state is provided by a State Estimation (SE) algorithm at high rates:
\begin{enumerate}
    \item The real-time control requires a convex grid model that can be solved efficiently. The SCs allow to reformulate the non-convex PF model into a linear constraint where the uniqueness of the optimal solution is guaranteed. \cite{christakou2015real}
    \item The closed-loop formulation of the SC requires only knowledge of the state of the grid and its admittance matrix. 
    In this paper, the states are estimated multiple times per second with very low latency by the linear SE that simultaneously computes the state of the AC and DC grids \cite{willem_SE}. Because two consecutive states are not changing significantly, the linear approximation is valid and allows for a very efficient computation with almost no loss in accuracy. 
\end{enumerate}




The SC-based optimal control is experimentally validated on the hybrid AC/DC microgrid developed at the EPFL. The hybrid network consists of 18 AC nodes, 8 DC nodes, and 4 converters that interface the AC and DC systems at different nodes. The four ICs operate in voltage control mode; that is, the DC voltage and the reactive power are regulated. Furthermore, resonant DC/DC converters are present to regulate the power flow in the DC network. Various resources are connected to the AC grid, such as three photovoltaic plants, an electric vehicle charging station (EVCS), and a controllable load that acts as a household. \\


The structure of the paper is as follows: Section \ref{Sec:Methodology} discusses the analytical computation of the SC and presents the formulation of the optimal control problem. In Section \ref{Sec:ExperimentalSetup}, the hybrid AC/DC grid and its resources are presented. Section \ref{Sec:ExperimentalValidation} presents the results of the experimental validation.

\section{Methodology}
\label{Sec:Methodology}

\subsection{Hybrid AC/DC network model}

\begin{figure}[!h]
\centering
  \includegraphics[width=0.99\linewidth]{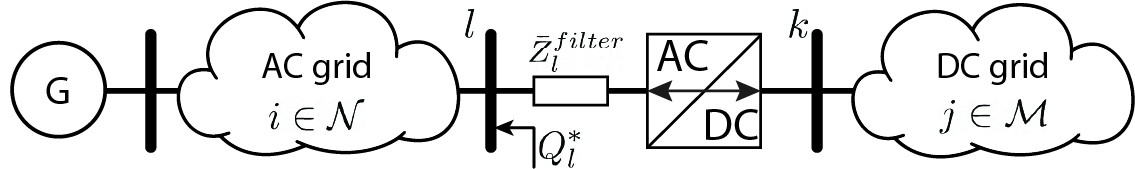}
  \caption{The generic hybrid AC/DC network. Only one AC/DC converter is displayed for simplicity.}
  \label{gengrid}
\end{figure}

Consider a generic hybrid AC/DC network with $i \in \mathcal{N}$ AC nodes and $j \in \mathcal{M}$ DC nodes, where buses $(l, k) \in \Gamma$ are the couples of AC/DC converter buses (see Fig.\ref{gengrid}). Furthermore, we assume $l \in \mathcal{N} $ and $k \in \mathcal{M}$. 

The AC network consists of three types of buses: a slack node $(\mathcal{N_{\text{slack}}})$, \textit{PV} nodes $(\mathcal{N_{PV}})$ and \textit{PQ} nodes $(\mathcal{N_{PQ}})$, and is modelled using the standard PF theory. The AC network is described as $\overbar{\mathbf{I}}^{ac} = \overbar{\mathbf{Y}}^{ac} \overbar{\mathbf{E}}^{ac} $, where $\overbar{\mathbf{E}}^{ac}$ is the phase-to-ground nodal voltage vector, $\overbar{\mathbf{I}}^{ac}$ the nodal current injections and $\overbar{\mathbf{Y}}^{ac}$ the compound admittance matrix, which is assumed to be known. 

The DC network is modelled identically to the AC network with $Q = 0$ and $\overbar{Z} = R$ to reuse the AC-PF theory. There are two types of nodes in the DC grid: voltage controllable nodes: \textit{V} nodes $(\mathcal{M_{V}})$, and power controllable nodes: \textit{P} nodes $(\mathcal{M_{P}})$. The DC network is described as $\mathbf{I}^{dc} = \mathbf{Y}^{dc} \mathbf{E}^{dc} $, with $\mathbf{E}^{dc}$ the DC voltage and $\mathbf{I}^{dc}$ the DC nodal current injections.  $\mathbf{Y}^{dc}$ is the compound admittance matrix of the DC grid.

Therefore, $\mathcal{N} = \mathcal{N}_{slack} \cup \ \mathcal{N}_{PQ} \cup \ \mathcal{N}_{PV} \cup \ \Gamma_l $ and $\mathcal{M} = \mathcal{M}_{P_{dc}} \cup \ \mathcal{M}_{V_{dc}} \cup \ \Gamma_k $

The AC and DC networks are interconnected by one or more ICs (that is, $\lvert \Gamma \rvert \geq 1$) and can operate under different control modes. 
Because of the nature of the ICs, which are typically Voltage Source Converters (VSC), it is not possible anymore to use the traditional PF theory, and an extension is needed where the model equations are dependent on the converter's operational mode: $E_{dc}-Q_{ac}$ or $P_{ac}-Q_{ac}$ nodes. 
Tab. \ref{NodeTypes} gives an overview of the possible node types in hybrid AC/DC grids. Note that at least one IC or DC source is required to impose the DC voltage ($E_{dc}$) \cite{baradar2011modeling}. The generic and unified power flow model for hybrid AC/DC networks is presented in \cite{willem_PF} and used in the experimental validation of the optimal control algorithm in this work.

\renewcommand{\arraystretch}{1.5}
\begin{table}[!h]
\caption{Different types of nodes in hybrid AC/DC networks and their known and unknown variables.}
\resizebox{1.05\columnwidth}{!}{%
\begin{tabular}{lllll}
\textbf{Bus Type   }             & \textbf{IC contrl. }                     & \textbf{Known var. }                                             & \textbf{Unknown var. } & \textbf{Index}\\ \hline
AC slack                   &                                  & $\lvert E_{ac} \rvert$, $\angle E_{ac}$                               & $P_{ac}$,$Q_{ac}$       & $s \in \mathcal{N}_{slack}$   \\ \hline
$P_{ac}$, $Q_{ac}$                 &                                  & $P_{ac}$,$Q_{ac}$                                                & $\lvert E_{ac} \rvert$, $\angle E_{ac}$    & $i \in \mathcal{N}_{PQ}$            \\ \hline
$P_{ac}$, $\lvert E_{ac} \rvert$                 &                                  & $P_{ac}$,$\lvert E_{ac} \rvert$                                                & $Q_{ac} $, $\angle E_{ac}$  & $i \in \mathcal{N}_{PV}$              \\ \hline
\multirow{2}{*}{$IC_{ac}$} & $P_{ac}$ - $Q_{ac} $                & $P_{ac}$ $Q_{ac}$     & $\lvert E_{ac} \rvert$, $\angle E_{ac}$           &  $l \in \Gamma_{PQ}$  \\ \cdashline{2-5}
                        & $E_{dc}$ - $Q_{ac}  $                   & $Q_{ac}$              & $P_{ac}$ $\lvert E_{ac} \rvert$ $\angle E_{ac}$     &  $l \in \Gamma_{E_{dc}Q}$        \\  \hline
\multirow{2}{*}{$IC_{dc}$} & $P_{ac}$ - $Q_{ac}$                 & $P_{dc}$              & $E_{dc}$           &  $k \in \Gamma_{PQ}$   \\ \cdashline{2-5}
                        & $E_{dc}$ - $Q_{ac} $                    & $E_{dc}$              & $P_{dc}$           &   $k \in \Gamma_{E_{dc}Q}$  \\  \hline
$P_{dc}$                     &                                  & $P_{dc}$                                                    & $E_{dc}$          & $j \in \mathcal{M}_{P}$    \\ \hline
$E_{dc}$                     &                                  & $E_{dc}$                                                    & $P_{dc}$           & $j \in \mathcal{M}_{V}$   \\ \hline
\end{tabular} \label{NodeTypes}
}
\end{table}

\subsection{Analytical computation of the sensitivity coefficients}

The PF equations of hybrid AC/DC networks are strongly non-linear \cite{willem_PF}. As discussed in Section \ref{sec:intro}, the PF model can be linearised around its operating point to include it in the OPF formulation. This allows for a more efficient computation with almost no loss of accuracy. Especially in real-time control where the states are computed with sub-second time resolution, consecutive states will not vary much. Therefore, linearising the grid constraints around its current operating point is a very good approximation for the next timestep. 

The linearised grid model, i.e., sensitivity coefficients, is computed in an analytical way as presented in \cite{willem_SC} and briefly discussed here. 

In a generic case, the set of controllable variables $\mathcal{X}$ consists of \eqref{eq:set_X}
\begin{align}
&\mathcal{X} = \left\{ P^{\ast}_{i},Q^{\ast}_{i},\lvert \overbar{E}_{i}  \rvert ^{\ast} ,P^{\ast}_{j},E^{\ast}_{j},P^{\ast}_l,Q^{\ast}_l , {E}_{k}^{\ast} \right\} \nonumber \\
&\qquad \qquad \qquad \qquad \qquad \quad \forall \ i \in \mathcal{N}, \ j \in \mathcal{M}, \ (l,k) \in \Gamma\label{eq:set_X}
\end{align}
where, following the conventions made in \ref{Sec:Methodology}, $P^{\ast}_i$, $Q^{\ast}_i$ and $\lvert \overbar{E}_{i}  \rvert ^{\ast}$ represent the PQ and PV nodes in the AC grid, $P^{\ast}_j$, and $E^{\ast}_j$ represent the P and V nodes in the DC grid, and $P^{\ast}_l$, $Q^{\ast}_l$ and $E^{\ast}_k$ represent the setpoints of the IC.

The closed-form analytical expression of the voltage SC is computed by taking the partial derivative of the PF equations with respect to the controllable variables $\mathbf{\mathcal{X}}$. The partial derivatives of the load flow model are shown in Appendix \ref{sec:appendix} in \eqref{eq:SC_model}. Next, by regrouping the terms in \eqref{eq:SC_model}, a linear system of equations is obtained:
\begin{eqnarray}
   \mathbf{A} \mathbf{x}(\mathcal{X}) = \mathbf{u}(\mathcal{X}).
\end{eqnarray}

$\mathbf{x}(\mathcal{X})$ is the vector of partial derivatives of the AC, DC and IC nodal voltages with respect to $\mathcal{X}$ \eqref{eq:X} \footnote{The last two sets of partial derivatives of \eqref{eq:X} represent only real voltages since these variables refer to the DC system}. 
\begin{align}
    &\mathbf{x}(\mathcal{X}) =  \left[  
    \frac{\partial \lvert \overbar{E}_{i} \rvert } {\partial \mathcal{X}}, 
    \frac{\partial \angle \overbar{E}_{i}}{\partial \mathcal{X}}, 
    \frac{\partial \lvert \overbar{E}_{l} \rvert }{\partial \mathcal{X}}, 
    \frac{\partial \angle \overbar{E}_{l}}{\partial \mathcal{X}} , 
    \frac{\partial {E}_{k} } {\partial \mathcal{X}}, 
    \frac{\partial {E}_{j} } {\partial \mathcal{X}} \right] \nonumber \\
    &\qquad \qquad \qquad \qquad \qquad \quad \forall \ i \in \mathcal{N}, \ j \in \mathcal{M}, \ (l,k) \in \Gamma
    \label{eq:X}
\end{align}
The matrix $\mathbf{A}$ is identical for every controllable variable $P^{\ast}$, $Q^{\ast}$ or $E^{\ast}$ and, therefore, only has to be computed once. Furthermore, the system of equations only has to be solved for the controllable variables of our interest. Therefore, this closed-form analytical method is computationally more efficient than the traditional method involving the inverse of the Jacobian of the PF model.

The branch current sensitivity coefficients can be obtained by using the network admittance matrix. The AC and DC current flow between the lines $i$ and $n$ and $j$ and $m$ can be expressed as:
\begin{equation}
    \overbar{I}_{i,n} = \overbar{Y}_{i,n}^{ac} \left( \overbar{E}_{i} - \overbar{E}_{n} \right) \ \text{and} \ I_{j,m} = Y_{j,m}^{dc} \left( E_{j} - E_{m} \right)
\end{equation}
Therefore, the branch current sensitivity coefficients are described as \eqref{eq:SC_I}.

\begin{eqnarray}
     \frac{\partial \lvert {\overbar{I}_{i,n}} \rvert}{ \partial \mathcal{X}} = & \overbar{Y}_{i,n}^{ac} (\frac{\partial \lvert {\overbar{E}_{i}} \rvert}{ \partial \mathcal{X}} - \frac{\partial \lvert {\overbar{E}_{n}} \rvert}{ \partial \mathcal{X}}) \\
     \frac{\partial {I_{j,m}}}{ \partial \mathcal{X}} = & Y_{j,m}^{dc} (\frac{\partial {E_{j}}}{ \partial \mathcal{X}} - \frac{\partial {E_{m}}}{ \partial \mathcal{X}}) \label{eq:SC_I}
\end{eqnarray}

Next, using the expressions of the voltage and current SCs derived above, the grid constraints can be formulated as a matrix representation that can be easily used in the optimisation problem. Let $\mathbf{K}_{P}^{E,t}$ be the matrix of the voltage SCs with respect to the active power in the AC and DC network at timestep $t$ \eqref{SC_matrix}: 
\begin{equation}
   \mathbf{K}_{P}^{E}
   =   
   \left[ \begin{NiceMatrix} 
    \ddots &  &  & \ddots &  & \\   
    & \frac{\partial \lvert{\overbar{E}_{i}}\rvert }{ \partial \mathcal{P}_{i}} &  & 
    & \frac{\partial \lvert{\overbar{E}_{i}}\rvert }{ \partial \mathcal{P}_{j}} &  \\
    &  & \ddots & &  & \ddots \\
    \ddots &  &  & \ddots &  & \\   
    & \frac{\partial {{E}_{j}} }{ \partial \mathcal{P}_{i}} &  & 
    & \frac{\partial {E_{j}} }{ \partial \mathcal{P}_{j}} &  \\
    &  & \ddots & &  & \ddots \\
    \CodeAfter 
                \tikz \draw [dashed] (1-|4) -- (7-|4) ;
                \tikz \draw [dashed] (4-|1) -- (4-|7) ;
   \end{NiceMatrix} \right],
   \ \ \forall \ \substack{ \large i \in \mathcal{N},\\
                                \\
                     j \in \mathcal{M}} \label{SC_matrix}.
\end{equation}  
Every row represents the AC or DC voltage and every column the AC or DC active power injection.
The same analogy holds for the voltage SCs of the reactive power and the voltage: $\mathbf{K}_{Q}^{E,t}$ and $\mathbf{K}_{E}^{E,t}$, and the current SCs: $\mathbf{K}^{I,t}$. Note that $\mathbf{K}_{Q}^{E,t}$ has zero elements in its DC positions, as there is no reactive power in the DC grid.
Furthermore, the AC and DC networks are treated as one unified grid: $\overbar{\mathbf{E}} = [\overbar{\mathbf{E}}^{ac}, \mathbf{E}^{dc}]$, $\overbar{\mathbf{I}} = [\overbar{\mathbf{I}}^{ac}, \mathbf{I}^{dc}]$, $\mathbf{P} = [\mathbf{P}^{ac}, \mathbf{P}^{dc}]$ and $\mathbf{Q} = [\mathbf{Q}^{ac}, \mathbf{Q}^{dc}]$, where e.g. $\overbar{\mathbf{E}}$ represents the vector of all the nodal voltages.
Therefore, the compound admittance matrix of the complete hybrid AC/DC grid is written as $\overbar{\mathbf{Y}} = diag(\overbar{\mathbf{Y}}^{ac}, \mathbf{Y}^{dc})$.

Finally, the voltage and current constraints can be written as \eqref{sc_voltage} and \eqref{sc_current}, where $t$ indicates the timestep, $\Delta {\mathbf{P}}^{t} =  \mathbf{P}^{t} - \mathbf{P}^{t-1}$, $ \Delta \mathbf{Q}^{t} = \mathbf{Q}^{t} - \mathbf{Q}^{t-1}$ and $ \Delta \lvert \overbar{\mathbf{E}}^{t} \rvert  = \lvert \overbar{\mathbf{E}}^{t} \rvert  - \lvert \overbar{\mathbf{E}}^{t-1} \rvert $.

\begin{eqnarray}
    & \lvert \overbar{\mathbf{E}}^{t} \rvert = \mathbf{K}_{P}^{E,t} \Delta {\mathbf{P}}^{t} + \mathbf{K}_{Q}^{E,t} \Delta {\mathbf{Q}}^{t} + \mathbf{K}_{E}^{E,t} \Delta \lvert \overbar{\mathbf{E}}^{t} \rvert \label{sc_voltage} \\
    & \lvert \overbar{\mathbf{I}}^{t} \rvert \ = \mathbf{K}_{P}^{I,t} \Delta {\mathbf{P}}^{t} + \mathbf{K}_{Q}^{I,t} \Delta {\mathbf{Q}}^{t} + \mathbf{K}_{E}^{I,t} \Delta \lvert \overbar{\mathbf{E}}^{t} \rvert \label{sc_current}
\end{eqnarray}

The grid losses are also linearized following the same approach:
\begin{eqnarray}
    & {P}^{losses,t}  = \mathbf{K}_{P}^{P,t} \Delta \mathbf{P}^{t} + \mathbf{K}_{Q}^{P,t} \Delta \mathbf{Q}^{t} + \mathbf{K}_{E}^{P,t} \Delta \lvert \overbar{\mathbf{E}}^{t} \rvert , \label{sc_Ploss} \\
    & \ \ {Q}^{losses,t}  = \mathbf{K}_{P}^{Q,t} \Delta \mathbf{P}^{t} + \mathbf{K}_{Q}^{Q,t} \Delta \mathbf{Q}^{t} + \mathbf{K}_{E}^{Q,t} \Delta \lvert \overbar{\mathbf{E}}^{t} \rvert , \label{sc_Qloss}
\end{eqnarray}
where e.g. $\mathbf{K}_{P}^{P}$ is the vector of the partial derivatives of the grid losses with respect to the active power injections. 

Both the unified PF model and the analytical computation of the SCs of hybrid AC/DC networks are made publicly available to the interested reader on \url{https://github.com/DESL-EPFL} \cite{github}.

\subsection{Problem Formulation}

Without loss of generality, the objective of the control problem is to regulate the active and reactive power injections of the controllable resources and the power flows to and from the DC grid, such that the grid nodal voltage and branch currents are always within the permissible bounds. 
Therefore, the DC grid can, for instance, be used to redirect the power flow, in order to relax one or more of the grid constraints. At the same time, the ICs can also be used to inject/absorb reactive power, to e.g. control the power factor at the grid connection point (GCP) to satisfy the local grid code. 
We assume that the controllable resources are photovoltaic plants that can curtail their production: $P_{i}^{pv}, \ \forall \ i \in \mathcal{N}^{pv} $.

The objective we minimize at time $t$ is:
\begin{equation} 
\begin{aligned}\label{eq:OPF}
    \min_{ P_{i}^{pv,t}, Q_{l}^t, E_{k}^t  }    
    \big( Q_{s}^t \big)^2 + \sum_{i \in \mathcal{N}^{pv}} \big(  P_{i}^{pv,t} - \widehat{P}_{i}^{pv,t} \big)^2  + \big(  P^{losses,t} \big)^2 
\end{aligned}
\end{equation}


where the first term $Q_{s}^t$ minimises the reactive power injected into the upper layer grid (slack bus) and the second term minimises the losses and the PV curtailment. The third term represents the minimisation of the power losses and prevents the reactive power injected/absorbed by the ICs from counteracting each other. This could occur when the voltage SCs at the ICs nodes are very similar because the nodes are e.g. physically located close to each other.

The problem is solved with respect to the constraints:
\begin{subequations} \label{eq:OPF_con}
\begin{align} 
    & [ \mathbf{E}^{ac}_{min},  \mathbf{E}^{dc}_{min} ] \leq  \lvert \overbar{\mathbf{E}}^{t} \rvert \leq [ \mathbf{E}^{ac}_{max},  \mathbf{E}^{dc}_{max} ] \\
    & 0 \leq  \lvert \overbar{\mathbf{I}}^{t} \rvert \leq [ \mathbf{I}^{ac}_{max},  \mathbf{I}^{dc}_{max} ] \\
    & 0 \leq  P_{i}^{pv,t}  \leq \widehat{P}_{i}^{pv,t} && \forall \ i \in \mathcal{N}^{pv} \label{eq:OPF_con_pv}\\
    & -\widehat{P}_{l} \leq  P_{l}^t  \leq \widehat{P}_{l}, && \forall \ l \in \Gamma_{l} \label{eq:OPF_con_IC_P} \\
    & -\widehat{Q}_{l} \leq  Q_{l}^t  \leq \widehat{Q}_{l}, && \forall \ l \in \Gamma_{l} \label{eq:OPF_con_IC_Q} \\
    &Q_{s}^t = \sum_{i \in \mathcal{N} \cap \mathcal{N}_{slack}} Q_{i}^t + Q^{losses,t} \\
    & \eqref{sc_voltage}, \eqref{sc_current}, \eqref{sc_Ploss} \ \text{and} \ \eqref{sc_Qloss}
\end{align}
\end{subequations}

where \eqref{eq:OPF_con_pv} refers to the maximum photovoltaic generation that is computed using a short-term forecast. Constraints \eqref{eq:OPF_con_IC_P} and \eqref{eq:OPF_con_IC_Q} refer to the maximum active and reactive power limits of the IC.

\subsection{Real-time Control Architecture }

\begin{figure}[!h]
\centering
  \includegraphics[width=0.35\linewidth]{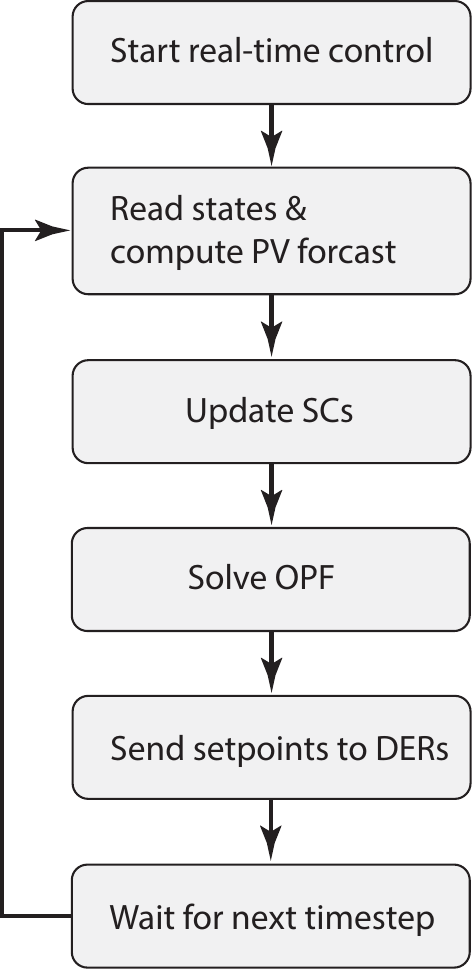}
  \caption{Flow-chart illustrating the real-time control architecture.}
  \label{RT_architecture}
\end{figure}

The real-time control architecture is shown in Fig. \ref{RT_architecture}. At each control timestep (every \SI{2}{s}), the state of the grid is provided by the state estimator \cite{willem_SE_exp}, and the GHI of the next timestep is used to calculate the Maximum Power Point (MPP) of the photovoltaics \cite{gupta2020grid}. Using the updated state estimate, the SCs are computed as described in Section \ref{Sec:Methodology} to represent the grid constraints in the optimisation problem. Next, the optimisation problem, described in \eqref{eq:OPF} and \eqref{eq:OPF_con}, is solved, and the optimal setpoints are sent to the controllable resources.

\section{Experimental Setup}
\label{Sec:ExperimentalSetup}

\subsection{Hybrid AC/DC grid}

\begin{figure*}
\begin{minipage}[c]{0.75\textwidth}
\includegraphics[width=\textwidth]{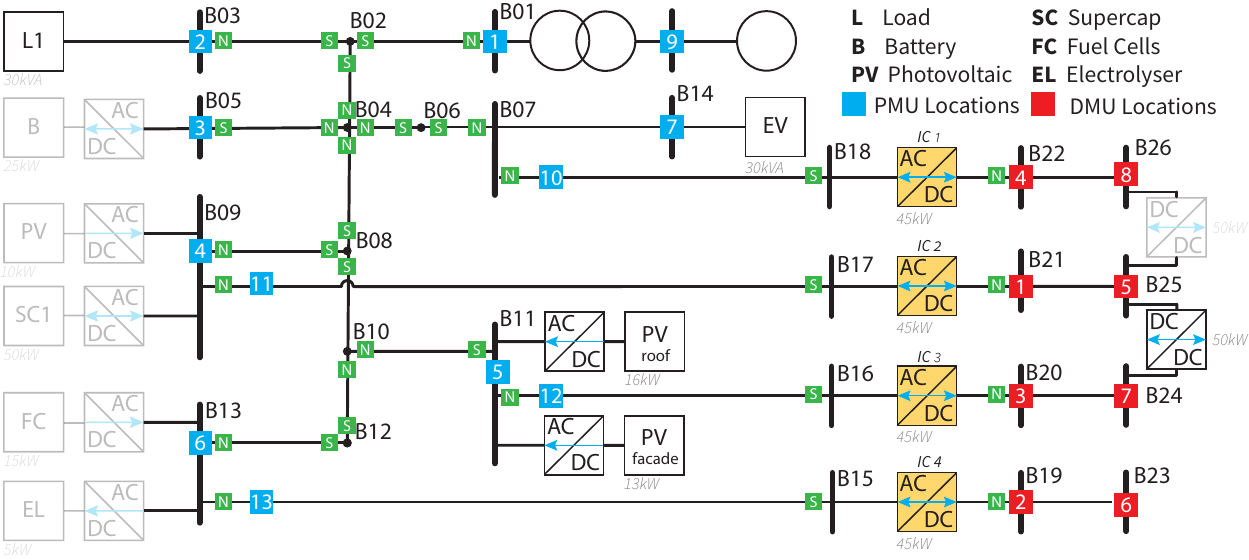}
\captionof{figure}{Hybrid AC/DC microgrid with the connected sources and loads, the maximum power rating is indicated. The table defines the boundary conditions of the simulation.}
\label{fig:Mgrid}
\end{minipage}
\begin{minipage}[c]{0.2\textwidth}
\footnotesize
    \vspace{-0.6cm}
    \begin{tabular}[b]{lll} 
       & \multicolumn{2}{c}{\textbf{AC}}                 \\  \cline{2-3} \cline{2-3}
       & \textbf{Bus Type}           & \textbf{Bus \#}   \\  \cline{2-3} \cline{2-3}
       & Slack (AC)                   & 1                 \\  \cline{2-3} 
       & PQ                          & 2-14              \\  \cline{2-3} \cline{2-3}
       & IC $(E_{dc} - Q_{ac} )$     & 15-18             \\  \cline{2-3} \cline{2-3}
       \vspace{-0.2cm}
       &                             &                   \\
       & \multicolumn{2}{c}{\textbf{DC}}                 \\  \cline{2-3} \cline{2-3}
       & \textbf{Bus Type}           & \textbf{Bus \#}   \\  \cline{2-3} \cline{2-3}
       & P                           & 24-26             \\  \cline{2-3} 
       & IC $(E_{dc} - Q_{ac}) $     & 19-22             \\  \cline{2-3} 
       \vspace{-0.1cm}
       &                             &                   \\
       & \multicolumn{2}{c}{\textbf{Resources}}                 \\  \cline{2-3} \cline{2-3}
       & \textbf{Type}              & \textbf{Rating}   \\  \cline{2-3} \cline{2-3}
       & PV roof                    & \SI{16}{kW}             \\  \cline{2-3} 
       & PV facade                  & \SI{13}{kW}             \\  \cline{2-3}
       & Load                       & \SI{30}{kVA}             \\  \cline{2-3}
       & IC $1-4$                   & \SI{45}{kVA}            \\  \cline{2-3}
       & DCT                       & \SI{30}{kW}            \\  \cline{2-3}
    \end{tabular}
  \label{gridtable} 
\end{minipage}
\vspace{-0.1cm}
\normalsize
\end{figure*}

The experimental validation is performed on the hybrid AC/DC grid located at the Distributed Electrical System Laboratory at the EPFL. The hybrid AC/DC microgrid consists of 18 AC nodes, 8 DC nodes and 4 ICs that operate under voltage control mode, that is, the pair ${E_{dc} - Q_{ac}}$ is controlled. Both grids have a base power of \SI{100}{\kilo VA} and a base voltage of \SI{400}{V_{ac}} and \SI{800}{V_{dc}}. The hybrid grid is connected to the medium voltage AC grid in the GCP at node $B01$.
The topology and parameters of the hybrid network are presented in Fig. \ref{fig:Mgrid}. 

\renewcommand{\arraystretch}{1.2}

The hybrid AC/DC grid hosts several \textbf{distributed resources}. At node $B03$, a controllable four-quadrant load of \SI{30}{kVA} is connected. The load is operated as a stochastic resource and represents the demand (active and reactive) of a typical household. The node $B14$ hosts an EVCS with a power rating of \SI{30}{kVA} that is also modelled as an uncontrollable resource. Three PV plants are connected at node $B11$. The first plant has a capacity of \SI{13}{kVA} and is mounted on the facade of a building. The commercial inverter does not allow for curtailment and will always track its MPP. The two other plants have a combined power rating of \SI{16}{kVA}, and its inverter allows for curtailment.

As shown in Fig. \ref{fig:Mgrid}, the DC lines are interconnected through \textbf{DC/DC converters}. The DC/DC converters are implemented as resonant DC transformers (DCT) and are based on open-loop control \cite{barcelos2022direct}. Therefore, they are not controlled through external setpoints, but the power transferred by the DCTs is proportional to the voltage difference between the primary and secondary side. Therefore, by using the IC to optimally regulate the DC voltage, a DC power flow is generated that can be used to redirect power in the AC grid.
This linear voltage-power relation is included in the optimisation problem as \eqref{eq:DCT}.
\begin{subequations} \label{eq:DCT}
\begin{eqnarray} 
    P_{1}^{DCT} = \ \ \alpha ( E_{1}^{DCT} - E_{2}^{DCT}) - P_{1, loss}\\
    P_{2}^{DCT} = -\alpha ( E_{1}^{DCT} - E_{2}^{DCT}) - P_{2, loss}
\end{eqnarray} 
\end{subequations}
where  $1$ refers to the primary and $2$ to the secondary side. The coefficient $\alpha$ equals $0.826 kW/V $. 
The two main contributions to the losses $P_{loss}$ are A) the magnetising current required to maintain the magnetising flux in the transformer's core and B) the equivalent ohmic losses of the DCT that are included in the DC grid model as an additional DC line with an equivalent series resistance of \SI{0.46}{\ohm}. 

It is worth mentioning that this is a simplified model of the DCT. The actual voltage-power profile is not fully linear and includes a zone around $\Delta E = 0$ where the operation is less stable and the power is close to zero. This inaccuracy in the model is the main cause of the uncertainty in the hybrid AC/DC model, as will be shown in Section \ref{Sec:ExperimentalValidation}.

\subsection{Sensing Infrastructure}

The computation of SCs requires knowledge of the nodal voltage at every bus of the hybrid AC/DC grid. 

Synchronised measurements are provided every \SI{20}{ms} by phasor measurement units (PMUs) for the AC grid and DC measurement units (DMUs) on the DC grid. The location of the PMUs and DMU is shown in Figure \ref{fig:Mgrid}. The PMUs are P-class devices and extract the phasors of measured currents and voltages using an enhanced interpolated discrete Fourier transform (e-IpDFT) \cite{romano2014enhanced}. The synchrophasor extraction is time synchronised using GPS and complies with the IEEE standard C37.118 \cite{C37118} with a total vector error of less than $0.14\%$. 
On the DC side, DMUs provide synchronised measurements of DC voltages and current injections. The DMUs use the same procedure as their AC variant, however, the e-IpDFT is replaced by an averaging block. The PMU and DMU measurements are streamed to the phasor data concentrate (PDC) using the user datagram protocol (UDP) \cite{asjaPDC}. The PDC time aligns the measurements with minimal latency and forwards them to the SE. 

The nodal voltage phasors are estimated by the SE and streamed to the real-time control every \SI{100}{ms}. The SE process for the hybrid AC/DC grid is described in detail in \cite{willem_SE_exp}. The method is based on a discrete Kalman filter and uses a unified and linear measurement model of the full hybrid grid, including the ICs. The state variables are typically the nodal voltage phasors. Because of the models' linear nature, the SE is able to compute the most likelihood state using the synchronised AC and DC measurements with subsecond time resolution.

\section{Experimental validation}
\label{Sec:ExperimentalValidation}

\begin{figure}[!h]
  \includegraphics[width=\linewidth]{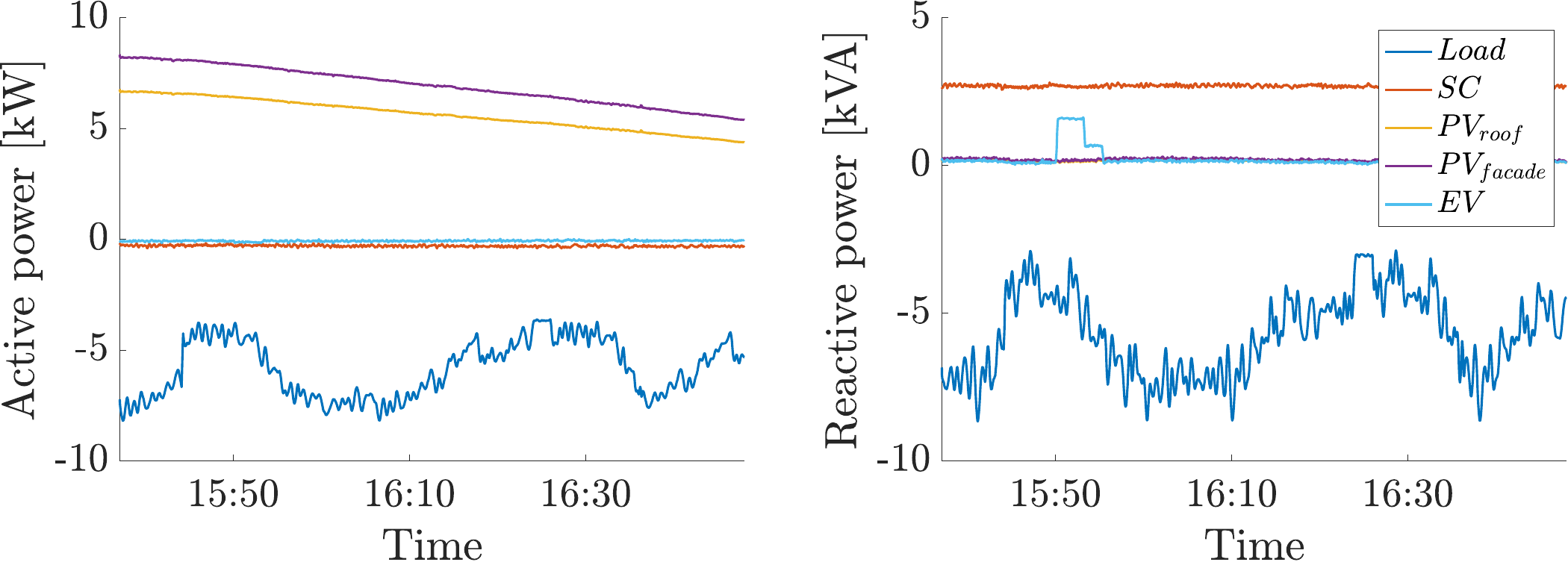}
  \caption{Active and reactive power of the DER.}
  \label{fig:DER_power}
\end{figure}

The closed-form analytical expression of the SCs is experimentally validated on the hybrid AC/DC microgrid described in Section \ref{Sec:ExperimentalSetup}. The experiment has been carried out for multiple days. However, in this paper, an interesting case is discussed that was conducted on September 15, 2023, between 15:37 and 16:48. The profiles of the active and reactive power injection from EVCS, supercapacitor, PV plants (MPP profile) and load emulator, which represents the demand of a typical household, are shown in Figure \ref{fig:DER_power}.   

\begin{figure}[!h]
  \includegraphics[width=\linewidth]{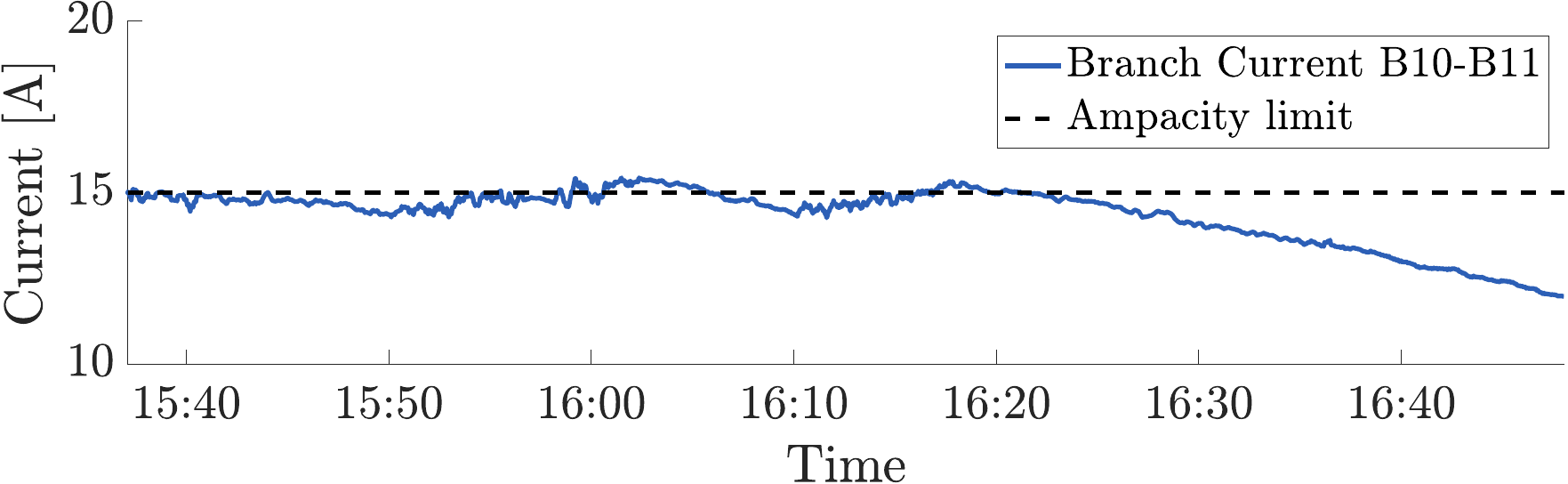}
  \caption{Current through line $B10-B11$ and its ampacity limit.}
  \label{fig:Amp_limits}
\end{figure}

\begin{figure}[!h]
  \includegraphics[width=\linewidth]{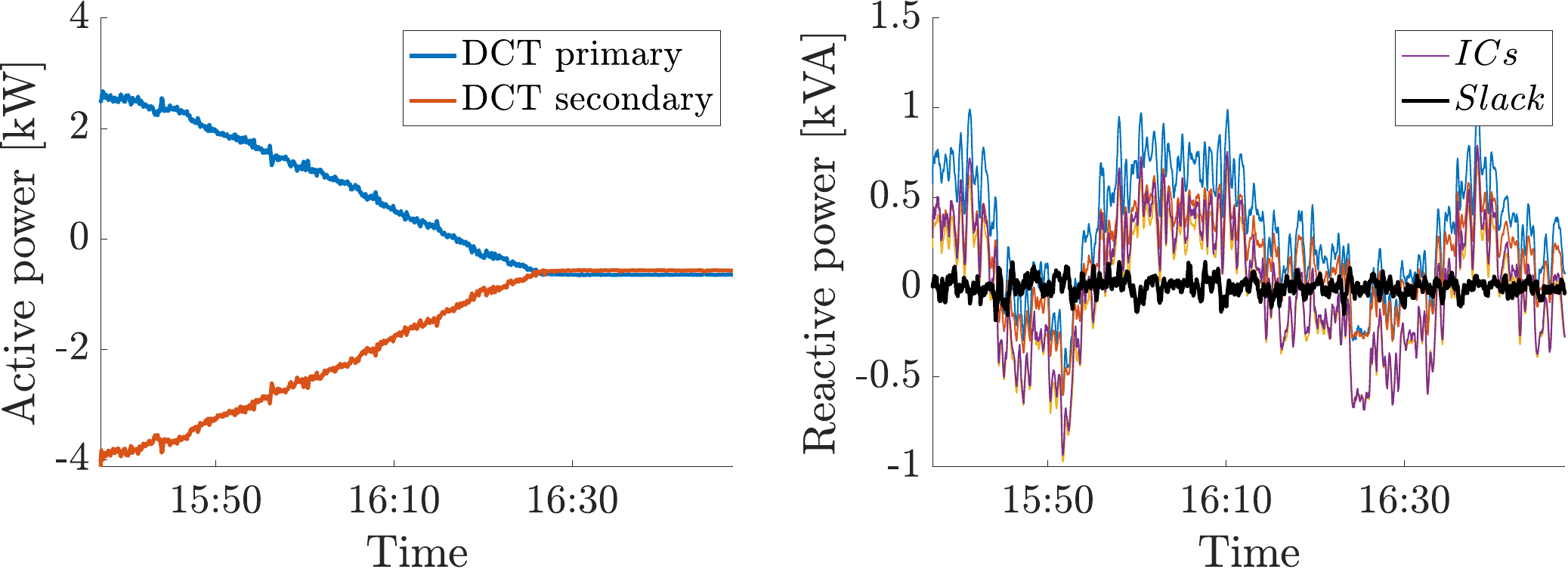}
  \caption{Active and reactive power of the controllable resources i.e., the ICs and DCT}
  \label{fig:CTRL_power}
\end{figure}

In Figure \ref{fig:DER_power}, it is shown that the PV generation in node $B11$ starts decreasing at \textit{15:37} from a cumulative \SI{14.87}{kW} to \SI{9.82}{kW} at \textit{16:48}. Therefore, the line $B10 - B11$ will gradually become less congested.
The ampacity limits of this line are \SI{15}{A}, and since we want to minimise the curtailment, a part of the photovoltaic production will be redirected through the DC grid. The branch current of the line $B10-B11$ and the ampacity limit are shown in Figure \eqref{fig:Amp_limits}. 
We can observe that the branch current slightly exceeds the ampacity limit in certain instances. The overshoot is never larger than \SI{0.35}{A}, corresponding to \SI{250}{W}. This is mainly due to the inaccuracy in the simplified DCT model, as already discussed in Section \ref{Sec:ExperimentalSetup}. The actual model deviates from a linear power-voltage relation for very small and very large powers. In further research, the DCT model will be better characterised and included in the real-time controller.
The optimal real-time control regulates the DC voltage of IC 2 and IC 3 to create a power flow through the DCT to avoid curtailment. The power through the DCT is shown in Figure \ref{fig:CTRL_power} (left). 
After 16:20, PV production no longer exceeds the line ampacity limit, and DCT stops transferring power. We also see that the power of the DCT does not go to zero after this time but reaches around \SI{-600}{W}. This is due to the transformer losses that are equally allocated to its primary and secondary side.

Additionally, the objective of real-time control aims at minimising the reactive power at the GCP. In Figure \ref{fig:CTRL_power} (right), the reactive power that is injected by the ICs is shown to minimise the reactive power at the slack node. We can see that the reactive power at the slack (in black) is very close to zero. Due to the loss term that is added to the objective in \eqref{eq:OPF}, the ICs only inject a minimal reactive power and do not counteract each other.

\subsection{Validation of the SC-based grid model}

\begin{figure}[H]
  \includegraphics[width=\linewidth]{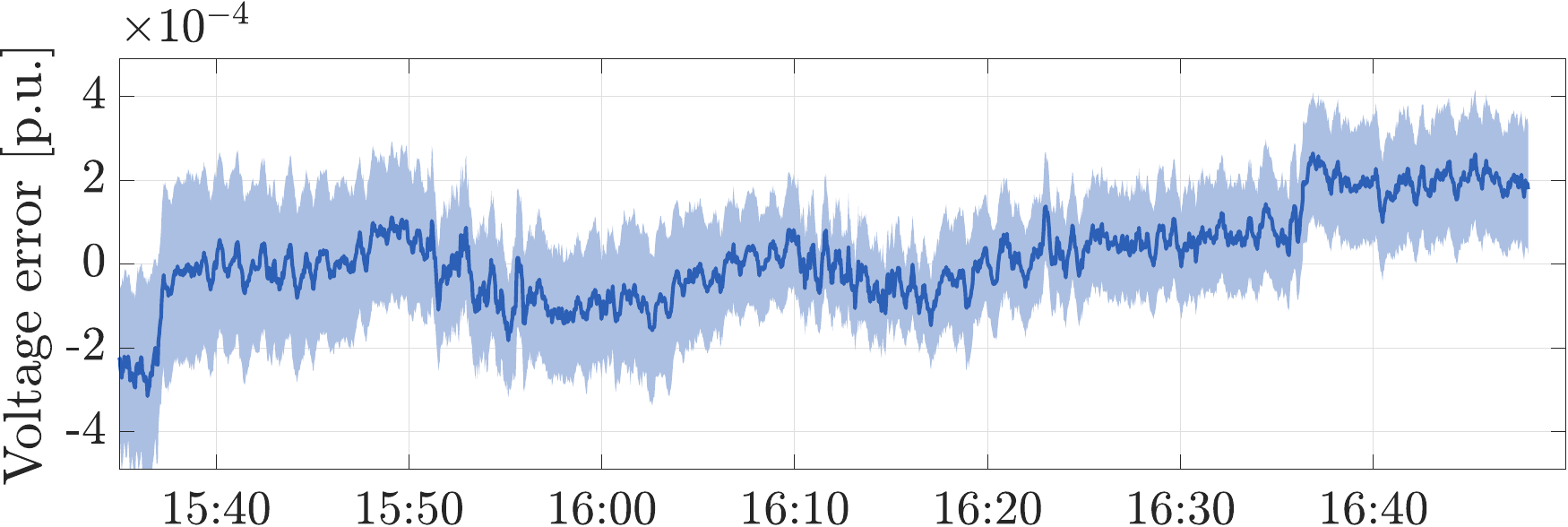}
  \caption{Voltage error of the SC-based model}
  \label{fig:accuracy}
\end{figure}

The accuracy of the grid model, which is represented by the SCs, is shown in Figure \ref{fig:accuracy}. The accuracy metric is defined as the difference between the actual grid voltage and the grid voltage that was expected by the control action. The mean of the voltage error of all the nodes is indicated in dark blue. The shaded light blue area represents the minimum and maximum voltage error at each timestep.
We see that the voltage error is very small with an average of \num{-7.62e-05} over the experiment. Therefore, the SC-based grid model is valid.

\subsection{Computational time analysis}
\begin{figure}[!h]
\centering
  \includegraphics[width=\linewidth]{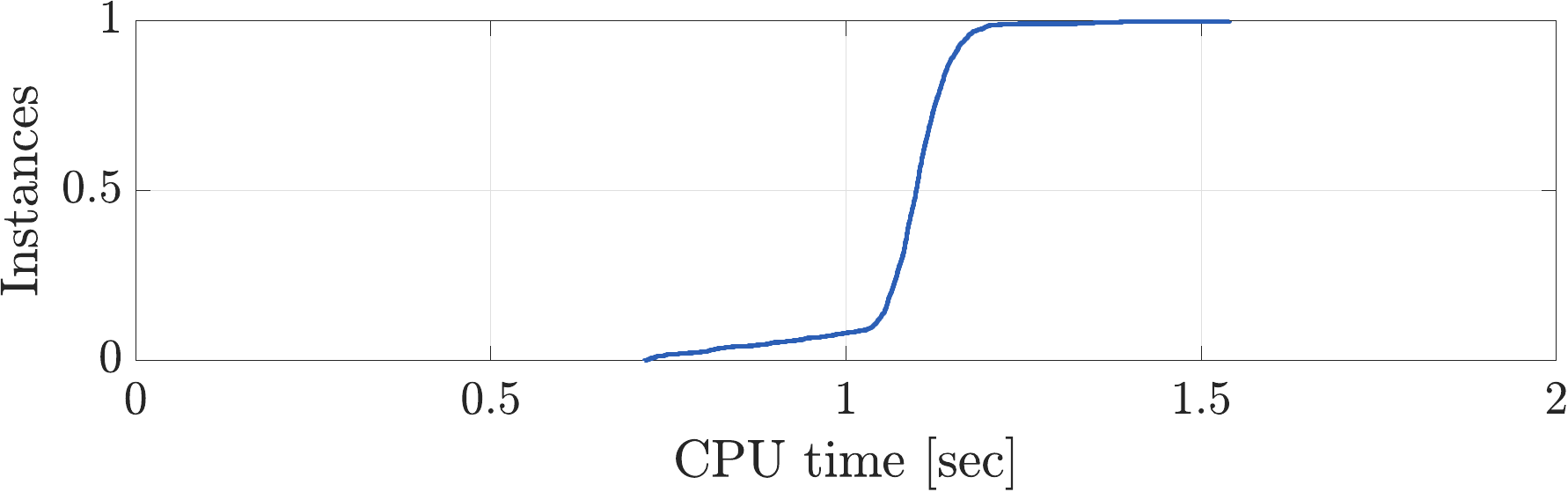}
  \caption{Cumulative distribution function of the computational time of the full real-time control.}
  \label{fig:CPU_time}
\end{figure}

The computational time of the real-time controller is evaluated and shown as a cumulative distribution function in Figure \ref{fig:CPU_time}. The CPU time includes the time of the full control process presented in Figure \ref{RT_architecture}, that is, fetching the grid state and GHI, computing the SC, solving the optimisation problem, and sending the setpoints to the resources. The overall time has an upper limit of \SI{1.5}{s} and is therefore very well suitable for critical real-time control processes.

\section{Conclusion}
\label{Sec:Conclusion}

In this paper, we present the experimental validation of a grid-aware optimal control of hybrid AC/DC microgrids leveraged by the closed-form computation of the voltage SCs. The analytical computation of the SC for hybrid AC/DC grids is presented in the author's previous work and is based on a unified PF model that accounts for the AC grid, DC grid, and the various operation modes of the ICs. 

The optimal SC-based control is validated on the hybrid AC/DC microgrid available at the EPFL. The microgrid consists of 18 AC nodes, 8 DC nodes, and 4 ICs. PMUs and DMUs provide synchronised AC and DC measurements to a linear SE that estimates the state of the network every \SI{100}{ms}. The state is streamed to the real-time control to update the SCs at every timestep. 

In a use case, it is shown that the real-time control operates correctly and avoids the need for PV curtailment by redirecting power through the DC grid to relax a congested AC line. The error in the SC-based model is on the order of $10^{-5}$ for the nodal voltages of both the AC and DC networks. Furthermore, it is shown that the full real-time control process (this includes reading the states, computing the SC, solving the OPF, and sending the updated resources setpoints) takes less than \SI{1.5}{s}. Therefore, it is very well applicable for the real-time control of hybrid AC/DC networks.

\appendix

\section{Sensitivity coefficient model}
\label{sec:appendix}

The voltage SC are computed as follows:
\begin{enumerate}
    \item Compute the partial derivative of the PF equations (presented in \cite{willem_PF}) to $\mathcal{X}$, shown in \eqref{eq:SC_model}.
    \item Regroup the partial derivatives in the form $A \mathbf{x}(\mathcal{X}) = \mathbf{u}(\mathcal{X})$, where $\mathbf{x}(\mathcal{X})$ are the voltage sensitivity coefficients $\frac{\partial \overbar{E}}{\partial \mathcal{X}}$ as described in \eqref{eq:X}.
    \item Solve the linear system of equations.
\end{enumerate}

To simplify the expressions of the partial derivative, two new variables are introduced $\overbar{F}_{i,n}^{ac} = \overbar{E}_i \underbar{Y}_{i,n}^{ac} \underbar{E}_n$ and ${F}_{j,m}^{dc} = {E}_j {Y}_{j,m}^{dc}{E}_m$. 

\vspace{3cm}

\begin{subequations}
\small
\allowdisplaybreaks
\begin{align}
        &\text{  \normalsize \textbf{AC nodes}}: \nonumber \\ 
        &\sum_{n \in \mathcal{N}} \Re \{ \overbar{F}_{i,n}^{ac} \} \left[ \frac{1}{ \lvert{\overbar{E}_{i} \rvert}} \frac{\partial  \lvert{\overbar{E}_{i} \rvert}}{ \partial \mathcal{X}} + \frac{1}{ \lvert{\overbar{E}_{n} \rvert}} \frac{\partial  \lvert{\overbar{E}_{n} \rvert}}{ \partial \mathcal{X}} \right] - \nonumber \\ 
        & \qquad \sum_{n \in \mathcal{N}} \Im \{ \overbar{F}_{i,n}^{ac} \} \left[ \frac{\partial \angle{\overbar{E}_{i}}}{ \partial \mathcal{X}} - \frac{\partial \angle{\overbar{E}_{n}}}{ \partial \mathcal{X}} \right] = \frac{\partial P_{i}^{\phi \ast}}{ \partial \mathcal{X}},  \nonumber \\
        & \qquad \qquad \qquad \qquad \qquad \qquad \qquad \qquad \quad \forall i \in \mathcal{N}_{PQ} \cup \mathcal{N}_{PV}  \label{PQ_p_sc}\\ 
        &\sum_{n \in \mathcal{N}} \Im \{ \overbar{F}_{i,n}^{ac} \} \left[ \frac{1}{ \lvert{\overbar{E}_{i} \rvert}} \frac{\partial  \lvert{\overbar{E}_{i} \rvert}}{ \partial \mathcal{X}} + \frac{1}{ \lvert{\overbar{E}_{n} \rvert}} \frac{\partial  \lvert{\overbar{E}_{n} \rvert}}{ \partial \mathcal{X}} \right] + \nonumber \\
        & \qquad \sum_{n \in \mathcal{N}} \Re \{ \overbar{F}_{i,n}^{ac} \} \left[ \frac{\partial \angle{\overbar{E}_{i}}}{ \partial \mathcal{X}} - \frac{\partial \angle{\overbar{E}_{n}}}{ \partial \mathcal{X}} \right] = \frac{\partial Q_{i}^{\phi \ast}}{ \partial \mathcal{X}}, \nonumber \\ 
        & \qquad \qquad \qquad \qquad \qquad \qquad \qquad \quad \qquad \forall i \in \mathcal{N}_{PQ} \label{PQ_q_sc}\\ 
        & \frac{\partial \overbar{E}_{i}}{\partial \mathcal{X}} = \frac{\partial \overbar{E}_{i}^{\phi \ast}}{\partial \mathcal{X}} , 
        \qquad  \qquad \quad \qquad \qquad \ \ \  \forall i \in \mathcal{N}_{PV} \label{PV_sc} \\ 
        \nonumber \\
        &\text{ \normalsize \textbf{DC nodes}}:  \nonumber \\ 
        & \sum_{m \in \mathcal{M}}  F_{j,m}^{dc}  \left[ \frac{1}{E_{j}} \frac{\partial E_{j}}{ \partial \mathcal{X}} + \frac{1}{E_{m}} \frac{\partial E_{m}}{ \partial \mathcal{X}} \right] 
        = \frac{\partial P_{j}^{\ast}}{ \partial \mathcal{X}}, \nonumber \\ 
        & \qquad \qquad \qquad \qquad \qquad \quad \qquad \qquad \qquad  \forall j \in \mathcal{M}_{P} \label{P_dc_sc}\\
        & \frac{\partial E_j}{\partial \mathcal{X}} = \frac{\partial E^{\ast}_{j}}{\partial \mathcal{X}}, 
        \qquad  \qquad \qquad \qquad \quad \qquad  \forall j \in \mathcal{M}_{V} \label{V_dc_sc} \\
        \nonumber \\ 
        &\text{  \normalsize \textbf{IC nodes}}: \nonumber \\ 
        & \sum_{n \in \mathcal{N}} \Re \{ \overbar{F}_{i,n}^{ac} \} \left[ \frac{1}{ \lvert{\overbar{E}_{i} \rvert}} \frac{\partial  \lvert{\overbar{E}_{i} \rvert}}{ \partial \mathcal{X}} + \frac{1}{ \lvert{\overbar{E}_{n} \rvert}} \frac{\partial  \lvert{\overbar{E}_{n} \rvert}}{ \partial \mathcal{X}} \right] - \nonumber \\
        &\sum_{n \in \mathcal{N}} \Im \{ \overbar{F}_{i,n}^{ac} \} \left[ \frac{\partial \angle{\overbar{E}_{i}}}{ \partial \mathcal{X}} - \frac{\partial \angle{\overbar{E}_{n}}}{ \partial \mathcal{X}} \right] - \frac{\partial P^{ loss}_{(l,k)}}{ \partial \mathcal{X}} = \frac{\partial P_{i}^{\ast}}{ \partial \mathcal{X}}, \nonumber \\ 
        & \qquad \qquad \qquad \qquad \quad \qquad \qquad \qquad \qquad  \forall l \in \Gamma_{PQ}  \label{EP_sc} \\ 
       & \sum_{n \in \mathcal{N}} \Im \{ \overbar{F}_{i,n}^{ac} \} \left[ \frac{1}{ \lvert{\overbar{E}_{i} \rvert}} \frac{\partial  \lvert{\overbar{E}_{i} \rvert}}{ \partial \mathcal{X}} + \frac{1}{ \lvert{\overbar{E}_{n} \rvert}} \frac{\partial  \lvert{\overbar{E}_{n} \rvert}}{ \partial \mathcal{X}} \right] + \nonumber \\ 
       & \sum_{n \in \mathcal{N}} \Re \{ \overbar{F}_{i,n}^{ac} \} \left[ \frac{\partial \angle{\overbar{E}_{i}}}{ \partial \mathcal{X}} - \frac{\partial \angle{\overbar{E}_{n}}}{ \partial \mathcal{X}} \right] -  \frac{\partial Q^{loss}_{(l,k)}}{ \partial \mathcal{X}}  = \frac{\partial Q_{i}^{\ast}}{ \partial \mathcal{X}} , \nonumber \\ 
        & \qquad \qquad \qquad \qquad \qquad \qquad \qquad \qquad \quad \forall l \in \Gamma_{PQ} \cup \Gamma_{V_{dc}Q} \label{EQ_sc} \\ 
       &\sum_{n \in \mathcal{N}} \Re \{ \overbar{F}_{l,n}^{ac} \} \left[ \frac{1}{ \lvert{\overbar{E}_{l} \rvert}} \frac{\partial  \lvert{\overbar{E}_{l} \rvert}}{ \partial \mathcal{X}} + \frac{1}{ \lvert{\overbar{E}_{n} \rvert}} \frac{\partial  \lvert{\overbar{E}_{n} \rvert}}{ \partial \mathcal{X}} \right] - \nonumber \\ 
       &  \sum_{n \in \mathcal{N}} \Im \{ \overbar{F}_{l,n}^{ac} \} \left[ \frac{\partial \angle{\overbar{E}_{l}}}{ \partial \mathcal{X}} - \frac{\partial \angle{\overbar{E}_{n}}}{ \partial \mathcal{X}} \right] + \frac{\partial P^{filter}_{(l,k)}}{ \partial \mathcal{X}} + \frac{\partial P^{loss}_{(l,k)}}{ \partial \mathcal{X}} = \nonumber \\ 
       & \sum_{m \in \mathcal{M}}  F_{k,m}^{dc} \left[ \frac{1}{E_{k}^{\ast}} \frac{\partial E_{k}^{\ast}}{ \partial \mathcal{X}} + \frac{1}{E_{m}} \frac{\partial E_{m}}{ \partial \mathcal{X}} \right], \nonumber \\ 
        & \qquad \qquad \qquad \qquad \qquad \qquad \qquad \qquad \quad  \forall (l,k) \in \Gamma_{V_{dc}Q}   \label{Edc_sc} 
\end{align} \label{eq:SC_model}
\normalsize
\end{subequations}

\vspace{3cm}

\bibliographystyle{IEEEtran}
\bibliography{IEEEexample.bib}

\end{document}